\documentclass[aps,superscriptaddress,showpacs,amsmath,amssymb,amsfonts,altaffillsymbol,twocolumn,notitlepage,prr]{revtex4-1}

\usepackage{float}
\usepackage{dcolumn}
\usepackage{bm}
\usepackage{hyperref}
\usepackage{xcolor}
\usepackage{tikz,pgfplots}
\pgfplotsset{compat=1.18}
\usepackage{ifthen}
\usepackage{graphicx}

\graphicspath{{./figures/}}

\begin{document}
\title{
Overlapping plastic events as a mechanism for irreversible dynamics\\ 
in amorphous solids under oscillatory shear
}

\date{\today}

\author{Asaf Szulc}
\affiliation{Department of Physics, Ben Gurion University of the Negev, Beer Sheva 84105, Israel}

\author{Ido Regev}
\email[Corresponding author: ]{regevid@bgu.ac.il}
\affiliation{
Department of Environmental Physics,
Jacob Blaustein Institutes for Desert Research, Ben-Gurion University of the Negev, Sede Boqer Campus 84990, Israel}
%
\begin{abstract}
The origin of the transition from asymptotically reversible to asymptotically irreversible response in amorphous solids subject to oscillatory shear is still unknown. It is known that the plastic events that result from shearing always involve localized particle rearrangements, but it is unclear why some are reversible while others are not. Here we show, using simulations and models, that overlaps between particle rearrangements caused by straining the solid in alternating directions can cause the response to become irreversible when they occur frequently. As the forcing amplitude increases, plastic events become more frequent, the number of such overlaps increases, and the probability of the system returning to previous states diminishes. 
\end{abstract}

\maketitle
%
\section{Introduction}
Since its discovery, the irreversibility transition in amorphous solids under oscillatory shear has attracted significant interest \cite{Regev2013,Keim2013,Fiocco2013,priezjev2013heterogeneous,Schreck2013,Schreck2013,Bhaumik2021,Leishangthem2017,reichhardt2023reversible,Kawasaki2016}.
This transition is observed when amorphous solids are subject to oscillatory shear
at athermal quasi-static (AQS) conditions for different strain amplitudes $\gamma_{\max}$. For small, sub-yield amplitudes, the system eventually reaches a periodic plastic state (a limit cycle in the terminology of dynamical systems), where plastic rearrangements repeat after each cycle and which stores an exact memory of the particle configurations that are part of the cycle\cite{Keim2019,Lavrentovich2017,Keim2020,Fiocco2015,Regev2015}. However, reaching these states requires many irreversible transient cycles, and when $\gamma_{\max}$ approaches a critical value $\gamma_c$, the number of cycles needed to reach a periodic state diverges. For $\gamma_{\max}>\gamma_c$, the system never reaches a limit cycle and is thus mechanically irreversible. This transition was demonstrated in experimental systems and atomistic simulations, where $\gamma_c$ was found to be close to or identical to the strain at which the material plastically yields (the yielding point) \cite{Regev2013,Keim2013,Fiocco2013,priezjev2013heterogeneous,Schreck2013}. 
Despite the large interest in the transition from reversible to irreversible dynamics and several attempts to explain it in terms of dynamics on a random energy landscape \cite{sastry2021models,mungan2021metastability,Szulc2020}, there is still no widespread agreement as to the nature and origin of this transition and these approaches do not provide a microscopic origin of irreversibility.

On the microscopic level, it is known that plastic events involve localized particle rearrangements known as {\it soft spots}~\cite{Argon1979,Falk1998,Maloney2006,Manning2011}. In two dimensions, a soft spot causes a displacement of four neighboring particles where two next-nearest neighbors become closer to each other and two nearest neighbors move further away from each other, resulting in a typical quadrupolar displacement field~\cite{Maloney2006,tyukodi2019avalanches}. The current understanding is that soft spots behave as two-level systems that can switch back and forth at specific stresses in response to external strains~\cite{Maloney2006,Mungan2019,Lindeman2021}. 
This point of view implies that most plastic transitions are reversible in the sense that there are strain paths that allow them to repeat. However, most plastic events are actually irreversible in the sense that they occur only once and can never repeat \cite{Lundberg2008,Mungan2019,Regev2021,Lindeman2021}. Nevertheless, the microscopic deformations in both cases seem qualitatively identical \cite{Lundberg2008,Mungan2019,Regev2021}. It is, therefore, not always clear what causes some events to be reversible while others irreversible.
\\

%
%
\begin{figure*}[tb]
\centering
\includegraphics[width=\linewidth]{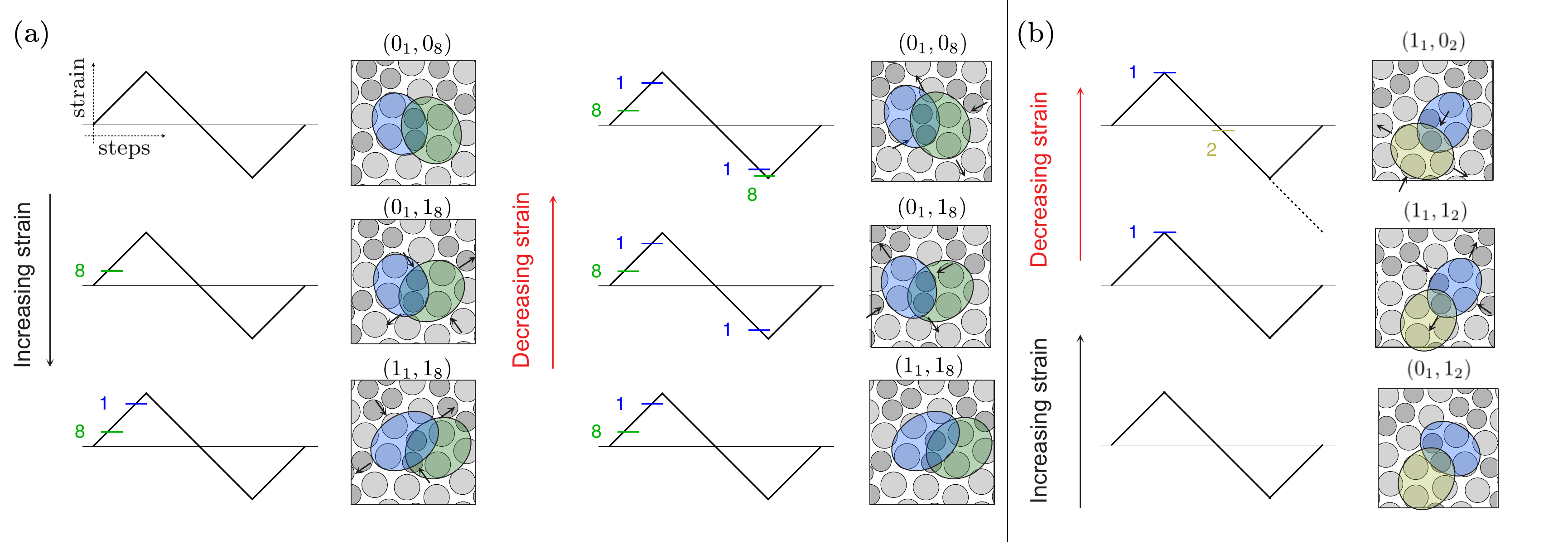}
\caption[Overlap examples in atomistic simulations]{
(a) A microscopic particle configuration exhibiting a regular overlap between two soft spots (regions marked by blue and green ellispses). 
(b) A microscopic particle configuration exhibiting an irregular overlap between two soft spots  (regions marked by blue and yellow ellipses). Both examples are taken from \cite{Szulc2022}. In the irregular configurations the coordinates were flipped $x_i\rightarrow -x_i$ such that a soft spot that was activated at a negative strain is now activated at a positive strain and vice-verse, in order to maintain consistency with the other figure. In both cases the strain at which a soft spot switches or switches back is shown as a horizontal line on a curve that shows the strain as a function of simulation steps (the equivalent of time for athermal quasistatic dynamics).
}
\label{Fig1}
\end{figure*}

Recent work on hysteresis and memory in amorphous solids has focused on modeling amorphous solids as a collection of interacting two-level systems representing soft spots~\cite{Lindeman2021,Szulc2022,Keim2021}. However, models with a fixed number of soft spots have a finite number of possible configurations, as will be explained below. Such models thus, have only transient irreversibility \cite{Lindeman2021} and eventually reach a limit cycle for any amplitude, and cannot exhibit yield under monotonic straining.
To address these issues, recent elastoplastic models chose to represent deformable regions as multi-level systems~\cite{khirallah2021yielding,Kumar2022, Liu2022}. This modeling strategy can allow for an infinite number of soft spots, and simulations of some of these models have demonstrated a diverging number of cycles at yield ~\cite{khirallah2021yielding,Kumar2022}. However, even in these models, the question of why the dynamics become irreversible at yield remains unanswered.

Here we propose that the sharing of particles, or overlaps, between soft spots is a plausible mechanism for the emergence of sustained irreversibility. By modifying and extending current models to include overlap effects observed in atomistic simulations, we show that overlaps caused by soft spots that switched due to alternating strain directions can prevent the system from reaching a limit cycle when they become frequent, which is the situation in the post-yield regime~\cite{Maloney2006,lin2014scaling,tyukodi2019avalanches}.\\\\\\

\section{Overlaps and their significance}

As mentioned above, a system comrpised of a finite number of hysterons will always reach a limit cycle.
To understand why, we note that in a system of $\mathcal{N}$ hysterons, where $\mathcal{N}$ is finite, a state of the system corresponds to a configuration of the possible states of the $\mathcal{N}$ hysterons. Each configuration of the system can thus be described in terms of a binary number with $\mathcal{N}$ digits of the form $s = 011000101\cdots010001$. In this number the value of the $n$th digit corresponds to the state of the $n$th hysteron which in the initial (reference) configuration can be either $0$, which indicates that it can switch due to an increase in the strain, or $1$, which indicates that it switches due to a decrease in the strain.
Note that out of all the possible $2^\mathcal{N}$ combinations of $0$s and $1$s, only a small subset represents configurations that are stable for some strain value, and that each stable configuration is stable for a range of strains/stresses (See discussion in \cite{Mungan2019b,Terzi2020,Mungan2019}). This means that for a fixed $\mathcal{N}$, the system can attain a finite number (significantly smaller than $\mathcal{N}$) of stable states. Since at zero temperature the dynamics is deterministic, once the system reaches a state that was visited previously, it continue to other states that were visited previously, and will therefore start periodic dynamics. Furthermore, since there is only a finite number of possible states, the system will always reach a state that was previously visited, eventually, and will thus always enter periodic dynamics (enter a limit cycle).

In an actual amorphous solid, the number of soft spots is not bounded. Since each soft spot is made of several particles (typically four), when a soft spot switches as part of a plastic event, some of its particles may belong to a soft spot that was switched previously. The new soft spot may thus disable previously switched soft spots from switching. 
Such sharing of particles between soft spots is certain to occur in finite systems at large enough strain amplitudes since if overlaps are prohibited, the space that was not yet subject to plastic deformation runs out eventually.
\begin{figure*}[tb]
\hspace*{-1cm} 
\includegraphics[width=0.8\linewidth]{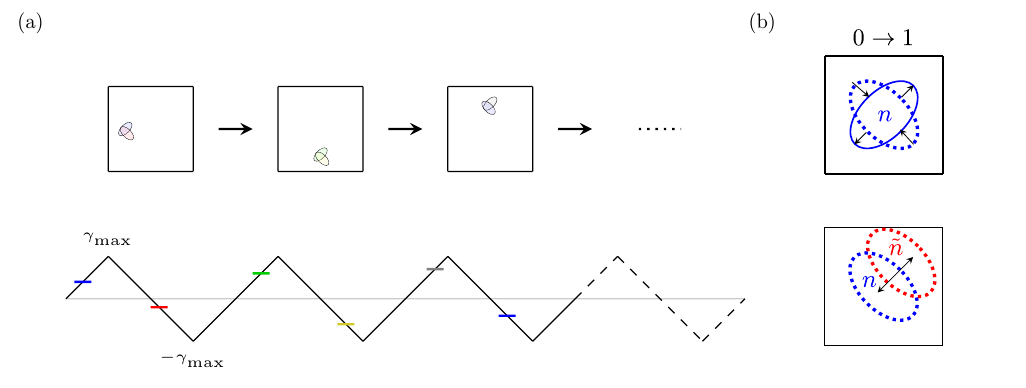}
\caption[]{
(a) At post-yield strain amplitudes $\gamma_{\max}$, deformation during one cycle involves many large avalanches occurring in both straining directions. This increases the chance that at least one irregular overlap will occur during each forcing cycle (i.e. after a deformation of the form $\gamma = 0\rightarrow \gamma_{\max} \rightarrow -\gamma_{\max} \rightarrow  0$) which will prevent the system from ever returning twice to the same zero-strain configuration, thus preventing the system from reaching a limit cycle. 
(b) Top: an illustration of the switching of a single soft spot from state $0$ (dotted) to state $1$ (continuous line) due to strain increase. 
Bottom: when irregular overlaps are allowed there is a probability $p_{ir}$ that a soft spot $\tilde n$ with $\sigma_{\tilde n}^+ < \sigma_{n}^+$ will switch before soft spot $n$ switches.}
\label{blow_out}
\end{figure*}

However, identifying such overlaps in experiments or atomistic simulations is challenging due to the difficulties in identifying individual soft spots, especially at large strain amplitudes when plastic events involve large avalanches that can include many different soft spots. In a previous work, we studied multiperiodic cycles observed in atomistic simulations of a system comprising 1024 particles subjected to AQS shear. By applying a special algorithm (which we developed) that employs a network representation of the dynamics, we identified all the soft spots switching during these cycles  \cite{Szulc2022}. The goal of this work was to understand the formation of multiperiodic cycles (i.e limit cycles in which the zero strain configuration repeats after more than one forcing cycle), as they involve both reversible and irreversible dynamics. As part of that work, we also identified overlaps between soft spots that were part of multiperiodic cycles. Nevertheless, the significance of overlaps to the dynamics was not clear and was thus not discussed in the manuscript.

A later examination of the aforementioned simulations revealed that overlaps come in two distinct scenarios. In the first scenario, two or more soft spots that switch at different strains overlap during a monotonic strain increase.
In most cases, when the strain is reversed, the soft spots switch in the opposite order as demonstrated in the example shown in Fig.\ref{Fig1}(a) (the exceptions will be discussed below). In Fig.\ref{Fig1}(a), soft spot 8 switches and following a further increase in the strain, soft spot 1 also switches. When the strain is decreased, soft spot 1 switches back and upon a further decrease in the strain, soft spot 8 also switches back. 
Thus, in this case, the switching exhibits a Last-In-First-Out (LIFO) ordering and the final zero strain state is identical to the zero strain state at the beginning of the cycle. 
In the second scenario, a soft spot that switches after the straining direction is reversed, disables a soft spot that switched due to an increase in the strain, as shown in the example in Fig. \ref{Fig1}(b). In this example soft spot 1 switches upon an increase in the strain while soft spot 2 switches upon a decrease in the strain before soft spot 1 is able to switch back (we have verified that soft spot 1 is not able to switch back even for significantly larger negative strains indicated by the dashed line). In the final zero-strain state, soft spot 1 and 2 are both switched and this state is thus not identical to the initial zero strain state. We refer to the first type of overlap as a {\it regular overlap}, as it involves sequential switching and switching back and the second type of overlap as an {\it irregular overlap} since it breaks the switching ordering.

The fact that the irregular overlap caused the system to not return to the initial zero strain state indicates that irregular overlaps can cause the system to not be able to reach a periodic cycle. To see this, consider a situation where the system is subject to $\gamma_{\max}$ that is larger than the yield strain. In this case, each shearing cycle involves a number of large avalanche events \cite{Nicolas2018} that occur due to straining in both positive and negative strains. One thus expects that the probability of irregular overlaps due to new plastic events will be large and that for large enough $\gamma_{\max}$, there will be at least one new such overlap during each forcing cycle, as is illustrated in Fig.~\ref{blow_out}(a). In this case, the system will not be able to repeat the same zero-strain configuration and periodic behavior will not be possible.  

\begin{figure}[tb]
\includegraphics[width=\linewidth]{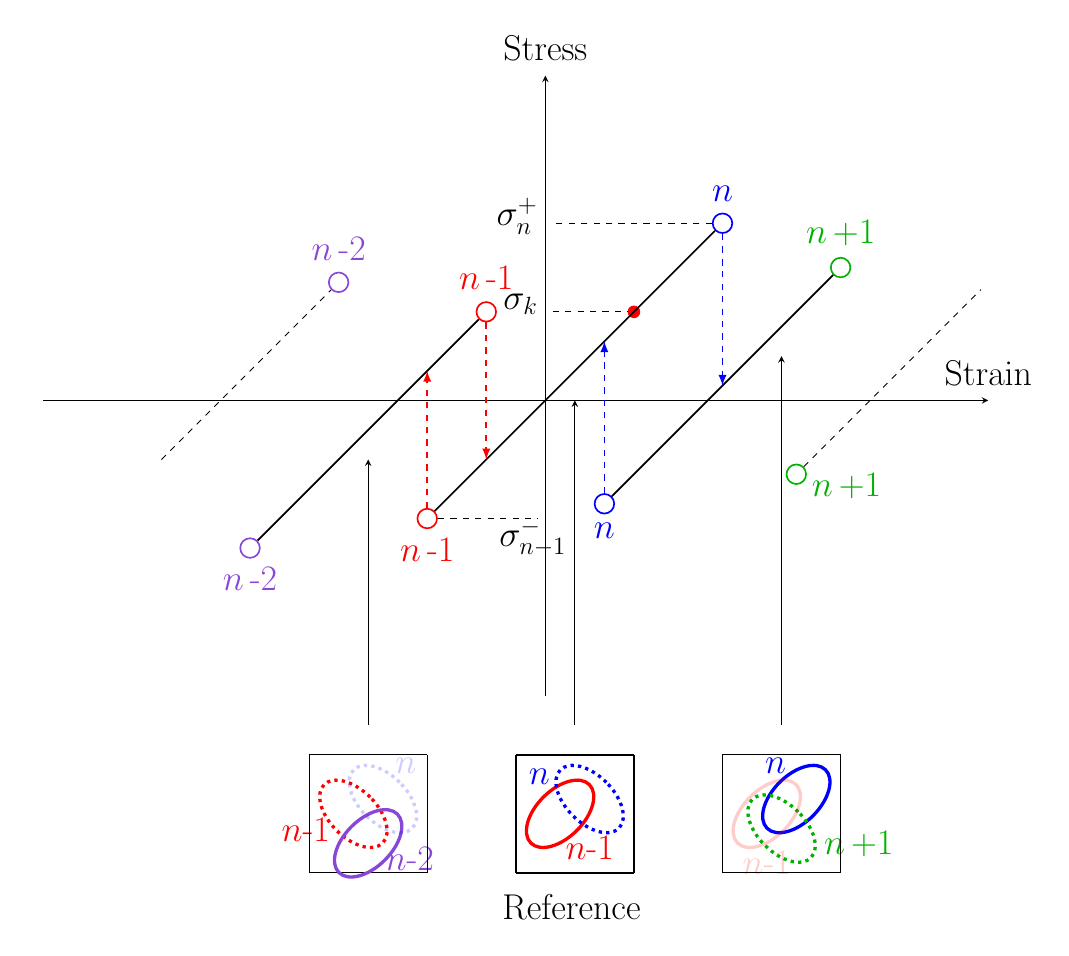}
\caption{
Illustration of the soft spot dynamics of a single site (site $k$ out of $N$ sites) in the model in the case where only regular overlaps are allowed.
At any given time, the site is in a particular configuration in which the stress is $\sigma_k$ and one soft spot
can switch due to a stress increase and another one can switch due to a stress decrease.
For example, in the highlighted reference configuration, soft spot n switches at $\sigma_{n}^+$ while soft spot $n-1$ switches
at $\sigma_{n-1}^{-}$ where $\sigma_{n}^- < \sigma_k < \sigma_{n}^+$.
If $\sigma_{n}^+$ is reached, the stress drops to a new value and then soft spot $n$ can switch back at $\sigma_{n}^-$ and soft spot $n+1$ can switch at $\sigma_{n+1}^+$. The former switch takes the system back to the previous (reference) configuration while the latter switch takes the system to a new configuration.}
\label{regular_il}
\end{figure}
\begin{figure}[tb]
\includegraphics[width=\linewidth]{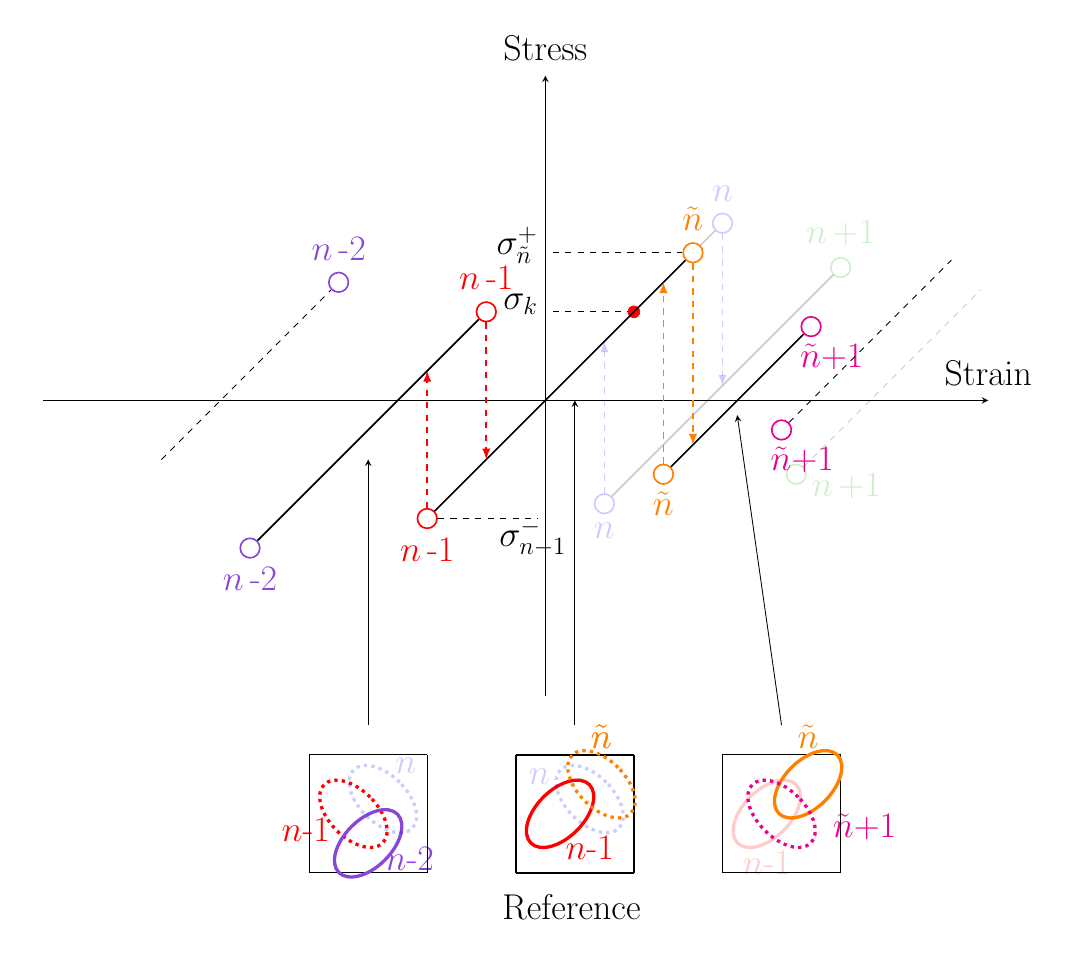}
\caption{
Illustration of the soft spot dynamics of the same site as in Fig.~\ref{regular_il} but where irregular overlaps are allowed.
For any site configuration, there is a probability $p_{ir}$ that a competing soft spot $\tilde n$ with $\sigma^+_{
\tilde n} < \sigma^+_n$ will switch before soft spot $n$ switches. In that case, upon a further increase (in the example shown in the illustration) in the strain, the system will reach configurations that were not reachable with purely regular overlaps (faded part of the illustration). However, if the strain is decreased (assuming no interactions with neighboring sites) the system can return to the configurations that were 
reachable by regular overlaps, assuming that no new irregular overlap occurs.}
\label{irregular_il}
\end{figure}
\begin{figure}[tb]
\centering
\includegraphics[width=\linewidth]{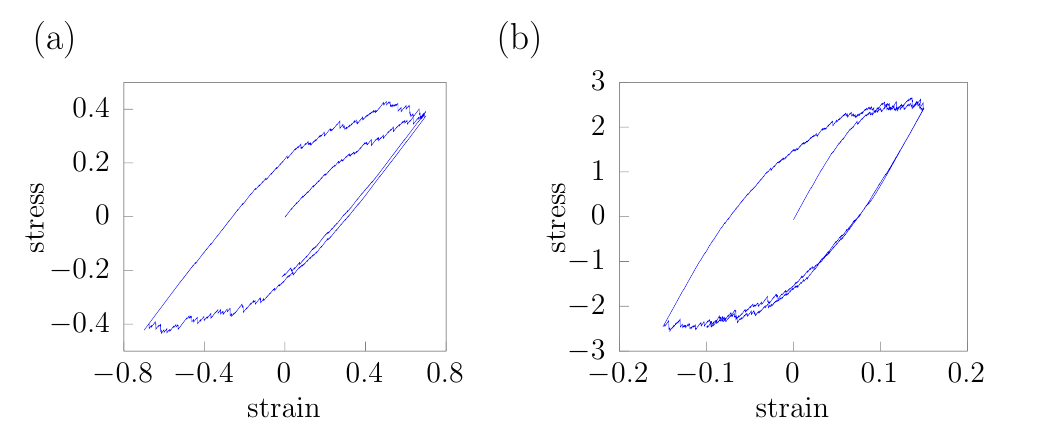}
\caption[Indications of irreversible dynamics]{
Comparison of the hysteresis loops generated by the model to hysteresis curves obtained from molecular dynamics simulations:
(a) stress-strain curve generated using the model for forcing amplitude $\gamma_{\max}=0.7$ with $p_{ir}=0.01$.
(b) stress-strain curve generated from atomistic simulations for forcing amplitude $\gamma_{\max}=0.15$ (data taken from the simulation results discussed in \cite{Regev2013}).
}
\label{irregular_irrev}
\end{figure}

\section{Model}
To better understand the contribution of overlaps to irreversibility, we developed a model of soft-spot dynamics that includes both regular and irregular overlaps.
In our model, soft spots are represented by interacting, hysteretic, two-state elements (hysterons), which we and others have used before to model the response for small strain amplitudes~\cite{Szulc2022, Lindeman2021, Keim2021, Bense2021, Hecke2021}. However, contrary to our previous models that used the strain as a state variable, here the stress in each hysteron is the primary variable, and switching is stress-activated, as is the case in other models of plasticity \cite{Hebraud1998,Bocquet2009,tyukodi2019avalanches,Kumar2022,Lin2014}. 
The interaction strength between soft spots is of the mean-field type with random, frustrated, coupling, similar to what we used in a previous work~\cite{Szulc2022} (see detailed description in the Supplementary material).
The soft spots are distributed in $N$ sites, where a site $k$ that belongs to one of these $N$ sites, is associated with a distinct stress value $\sigma_k$ that is modified at each simulation step due to an applied strain step $\delta\gamma$ by $\delta \sigma = \mu \delta\gamma$. Here $\mu$ is the shear modulus, which is assumed to be a material constant. Each site represents several soft spots that share some of their particles, and thus when one of them switches, others become disabled, similar to the situation in Figs.~\ref{Fig1}(a,b). 

In the following we first develop a model that allows only regular overlaps and then extend it to also include irregular overlaps. 
When only regular overlaps are allowed, each site has two switchable soft spots for each configuration of the system.
In Fig.\ref{blow_out}(b,top) we show a schematic demonstrating the switching of a single soft spot $n$ between two possible states due to an increase in the stress. When the stress is increased to $\sigma_n^+$, the soft spot switches from state $0$ to state $1$. Upon a decrease of the stress to $\sigma_{n}^-$, the soft spot switches back to state $0$. A soft spot can also have an initial state $1$ and switch to state $0$ upon a decrease of the strain.

In Fig.\ref{regular_il} we show typical dynamics in one site when only regular overlaps are allowed.
We take the configuration in the bold square to be a typical configuration which we use as a reference configuration (or reference state) for this specific site.
In this configuration there are two switchable soft spots, $n$ and $n-1$. Soft spot $n$ starts at state $0$ and switches to state $1$ if the stress is increased to $\sigma_{n}^+$. Soft spot $n-1$ starts at state $1$ and switches to state $0$ if the stress is decreased to $\sigma_{n-1}^-$. When soft spot $n$ switches, it disables soft spot $n-1$ from switching due to an overlap. The site then reaches a new configuration (box to the right) and a new soft spot $n+1$ that can switch due to an increase in the stress is added to the site. Starting in this configuration, soft spot $n$ can switch back from state $1$ to state $0$ if the stress is decreased to $\sigma_{n}^-$ which then disables soft spot $n+1$ but releases soft spot $n-1$ and brings the site back to the reference configuration. Similarly, starting from the reference configuration and decreasing the stress to $\sigma_{n-1}^-$, soft spot $n-1$ switches from $1$ to $0$, disables soft spot $n$ and causes the creation of a new soft spot $n-2$ that can switch if the strain is decreased further. The dynamics obtained from this scheme is similar in nature to the single-cell multi-well dynamics used in recent elastoplastic models \cite{khirallah2021yielding, Kumar2022, Liu2022}.

We next modify the model to include the possibility of irregular overlaps. To allow for such overlaps, whenever the dynamics reaches a configuration which was not visited before, we add a new soft spot to the site with probability $p_{ir}$. This is illustrated in Fig.~\ref{blow_out}(b,bottom) where soft spot $\tilde n$ with a random $\sigma_{\tilde n}^+ < \sigma_{n}^+$ is added to the cell. 
This introduces irregular overlaps since it allows a soft spot to become overlapped and disabled before switching back to its initial state. When the system reaches a previously visited system-wide configuration, we do not allow further soft spot additions (i.e. we set $p_{ir}=0$). This is compatible with the deterministic nature of the dynamics in the atomistic simulations and guarantees that once a previously visited state is reached, the system will enter a limit cycle. 
Fig.~\ref{irregular_il} demonstrates how the dynamics is modified when an irregular overlap is added at the reference configuration. Starting from $\sigma_k$, the stress is increased and at $\sigma_{\tilde n}$ soft spot $\tilde n$ switches from $0$ to $1$ which then disables soft spots $n$ and $n-1$ and causes the creation of a new soft spot $\tilde n + 1$ that can switch upon a further increase in the stress. The system thus enters an alternative path in the configuration space where soft spots $n$, $n+1$, $n+2$, $...$ cannot switch and soft spots $\tilde n$, $\tilde n + 1$, $\tilde n + 2$, $...$ are switched instead. Upon a decrease of the strain, the system can, in principle switch  $n-1$, $n-2$, $n-3$, $...$ back to their $0$ states.
However, there is a finite probability that another irregular overlap will occur before any of these soft spots switch back, and in that case at least some of these soft spots will not switch back, and alternative soft spots will switch instead.
\begin{figure}[tb]
\hspace*{-1cm} 
\includegraphics[width=0.8\linewidth]{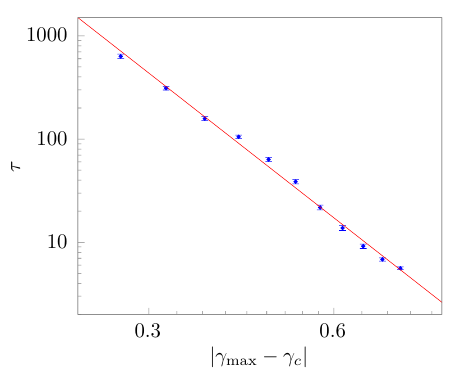}
\caption{
The average number of cycles $\tau$ needed to reach a periodic state as a function of \mbox{$|\gamma_{\max} - \gamma_c|$} for $p_{ir} = 0.01$. 
}
\label{power_law}
\end{figure}

\section{Results}

We first verifiy that the model reproduces the salient features of amorphous solids as observed in experiments and atomistic simulations. In Fig.~\ref{irregular_irrev} we compare the stress-strain curves obtained from the model \ref{irregular_irrev}(a) and atomistic simulations \ref{irregular_irrev}(b). It is clear that qualitatively, the behavior is very similar. In both cases there is observable hysteresis and in both cases large avalanches become more frequent in the post-yield regime. Other properties of the stress-strain response and avalanches are discussed in the supplementary material. 
To study the contribution of irregular overlaps to irreversibility, we compared simulations of the model with $p_{ir} = 0$ (strictly regular overlaps) and $p_{ir} = 0.01$ (a finite but small probability for irregular overlaps). In both cases, we simulated $100$ realizations of systems of $N = 250$ sites and studied the response to up to $2000$ forcing cycles. We found that in both cases the response to monotonically increasing strain was qualitatively identical (see Supplementary material), but the response to cyclic shear was extremely different:
when $p_{ir} = 0$ the system reaches a limit cycle for any forcing amplitude $\gamma_{\max}$ but when $p_{ir} = 0.01$ the number of cycles $\tau$, needed to reach a limit cycle seem to diverge at a critical amplitude $\gamma_c$ as shown in Fig.~\ref{power_law}. Fitting to $\tau \sim |\gamma-\gamma_c|^{\alpha}$ leads to a best fit for $\gamma_c = 0.97$ and $\alpha = 4.65(1)$. The exponent $\alpha$ obtained is larger than that observed in atomistic simulations, $\sim 2.6$, which could be attributed to using random, mean-field interactions rather than realistic quadrupolar interactions.
%
\begin{figure}[tb]
\includegraphics[width=\linewidth]{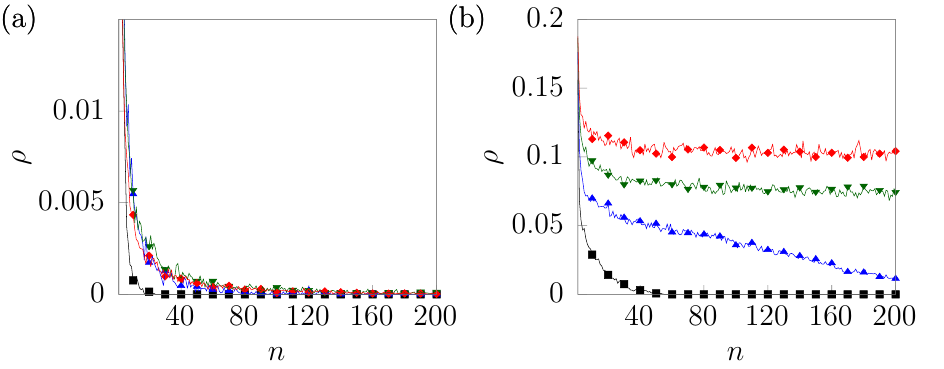}
\caption[Indications of irreversible dynamics]{
(a) The number of new soft spots $\rho$ that are added in each cycle as a function of the number of applied cycles $n$ for $p_{ir} = 0$. The same plot for $p_{ir} = 0.01$. The data is normalized such that $\rho$ is one at $n=0$ and $\gamma_{\max} = 0.4 (\bigcirc), 0.6 (\textcolor{blue}{\square}), 0.8 (\textcolor{green}{\triangle}), 1.2 (\textcolor{red}{\lozenge})$.
}
\label{Fig8}
\end{figure}

To better understand the difference between dynamics with strictly regular overlaps and dynamics that include irregular overlaps, we recorded the number of new soft spots added at each cycle $\rho$, which is an indication of the a-periodicity of the cycle. Fig.~\ref{Fig8}(a) shows that for the strictly regular $p_{ir}=0$, the number of new soft spots per cycle decreases rapidly to zero, irrespective of the driving amplitude $\gamma_{\max}$. However, when $p_{ir}=0.01$ (Fig.~\ref{Fig8}(b)), there is a transition as a function of $\gamma_{\max}$ between transient dynamics in which the number of new soft spots decays to zero and dynamics in which the number of new soft spots reaches a constant value. 
This is consistent with our understanding that an irreversible steady state requires the number of soft spots to become effectively infinite (note that $\rho=0$ does not mean that the system immediately enters a limit cycle, but similarly to a system with a fixed number of soft spots, it will eventually reach a limit cycle after a finite number of forcing cycles).

\begin{figure}[tb]
\includegraphics[width=\linewidth]{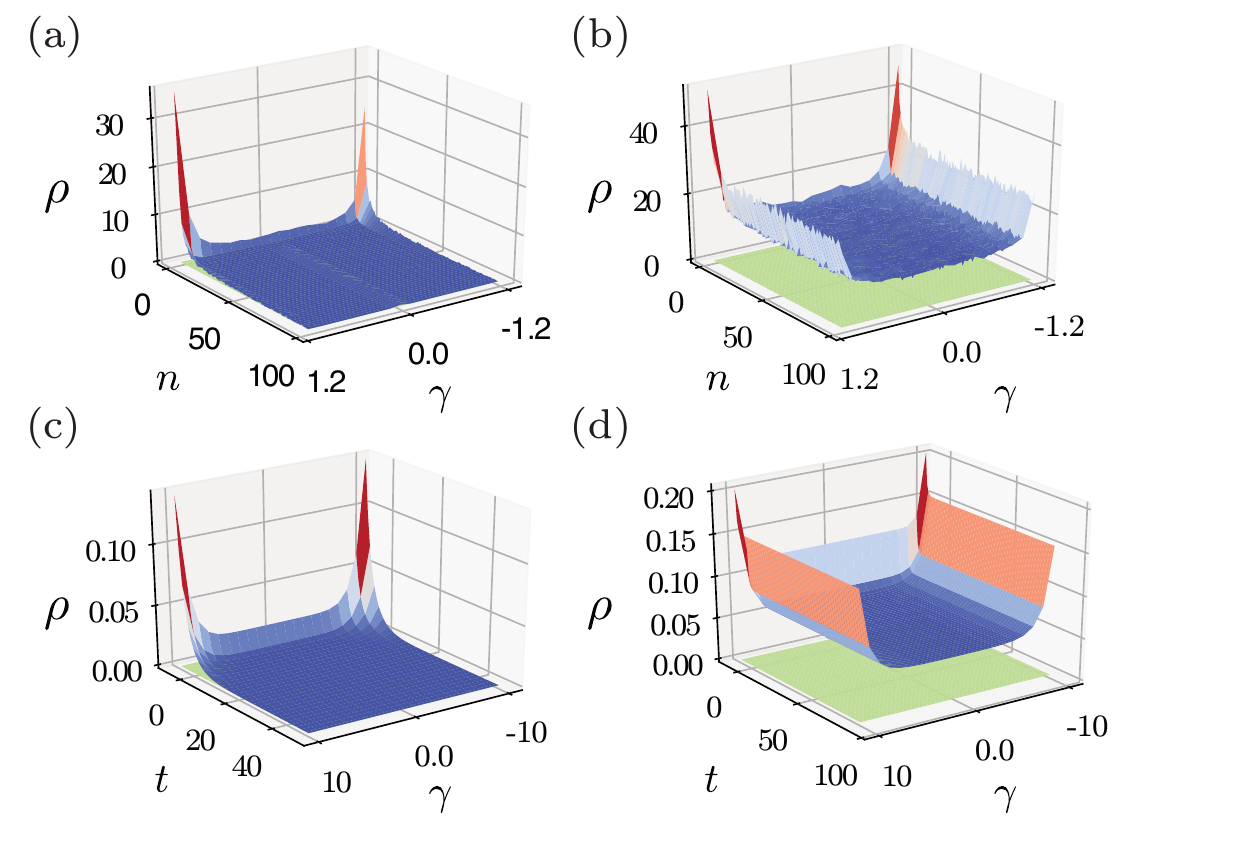}
\caption{The number of new soft spots added in each cycle $\rho$, as a function of the strain $\gamma$ and the cycle number $n$ starting from $n=2$, averaged over $100$ realizations for (a) $p_{ir} = 0$ and (b) $p_{ir} = 0.01$. Numerical solution of Eq.~(\ref{model}) as a function of $\gamma$ and $t$ with (c) $c=0$ and $r=-1$ (equivalent to regular dynamics) starting from $t=3$ and (d) $c=10$ and $r=0.5$ (equivalent to irregular dynamics) starting from $t=1$.
}  
\label{Fig4}
\end{figure}

The origin of the difference between dynamics that includes and excludes irregular overlaps can be understood by studying the number of new soft spots added per cycle $\rho$ as a function of the strain at which they are switched $\gamma$, and the number of cycles applied $n$. In Fig.~\ref{Fig4}(a,b), we show the average $\rho$ as a function of $n$ and $\gamma$ for a post-yield amplitude $\gamma_{\max}=1.2$, starting from the second cycle (we are interested in the transient distribution). When $p_{ir}=0$ (Fig.~\ref{Fig4}(a)), the number of new soft spots rapidly decays to zero for most of the strain range except for two ``boundary layers''  close to $\pm\gamma_{\max}$, and eventually, the number of new soft spots decays to zero in the entire strain-range. These boundary layers are also observed when $p_{ir}=0.01$ (Fig.~\ref{Fig4}(b)). However, in this case, the number of new soft spots per cycle reaches a non-zero steady state. The difference in the steady-state density between $p_{ir}=0$ and $p_{ir}>0$ stems from the fact that in the former, after the first forcing cycle, new soft spots are added only in the vicinity of the boundaries of the strain range (i.e., $\gamma = \pm\gamma_{\max}$) due to the hierarchical nature of regular overlaps, whereas in the latter, new soft spots are also added at intermediate strains due to the effect of irregular overlaps, which can occur at any strain value. This allows the number of soft spots added per cycle to reach a non-zero steady state in which new soft spots are continually added to the system.
%

To better understand this behavior, we developed a model based on a mean-field description of the dynamics, which is similar to approaches taken before to describe the irreversibility transition in dilute systems and the effect of jamming on such a transition \cite{ness2020absorbing,xu2013contact}.
The primary assumption is that the number of new soft spots added at the $n+1$ cycle depends on how many new soft spots were added at the $n$th cycle. If no new soft spots are added starting from some forcing cycle, the system eventually reaches a limit cycle, as explained above. 
The dynamical variable in our model is thus $\rho$, which represents the number of new soft spots added at a time $t$ on a domain $-\gamma_{\max} \leq \gamma \leq \gamma_{\max}$. The equation describing the change in $\rho$ is a nonlinear diffusion equation:
\begin{equation}
\rho_t = D\rho_{\gamma\gamma} + r\rho - c\rho^2\,,
\label{model}
\end{equation}
where the diffusive term $D\rho_{\gamma\gamma}$ takes into account random, mean field interactions as was shown in other works \cite{parley2020aging}. These lead soft spots to switch at different strains in the next cycle (the so-called scrambling \cite{Hecke2021}). The case $p_{ir}=0$ is modeled by taking $c=0$ and $r<0$, which accounts for the hierarchical property of regular overlaps that prevents the addition of new soft spots at intermediate strains, which then causes $\rho$ to degrade over time. 
For $p_{ir}>0$, we still assume that there is a degradation term proportional to $\rho$, but we also add a term that takes into account the addition of new soft spots at strains $-\gamma_{\max}<\gamma<\gamma_{\max}$ due to irregular overlaps. We assume that the rate of soft spot addition is proportional to the current number of new soft spots $\rho$, and the number of sites in which new soft spots were not added at time $t$. This term is thus of the form $\rho (N - \rho)$ where $N$ is the total number of sites in the system. Together these terms lead to the nonlinear part of Eq.~(\ref{model}).
Note that this is the simplest nonlinear term that leads to a transition between $\rho=0$ and a non-zero steady-state and that this term is equivalent in form to the mean-field theory of directed percolation \cite{henkel2008non}. 
To take into account the addition of new soft spots at $\pm\gamma_{\max}$ (for any $p_{ir}$), we assign nonlinear boundary conditions:
\begin{equation}
\rho_{\gamma}\Bigr|_{\substack{\gamma=\pm\gamma_{\max}}} = \pm[\rho - \rho^2]|_{\substack{\gamma=\pm\gamma_{\max}}}\,,
\end{equation}
where we have chosen the simplest boundary conditions that allow for steady states with either zero or non-zero $\rho$ starting from a non-zero $\rho$.  
In Fig.~\ref{Fig4}(c) we show numerical solutions of Eq.~(\ref{model}) for $D=0.75$, $c=0$ and $r=-1$ whereas in Fig.~\ref{Fig4}(d) we show the results for $D=0.75$, $c=10$ and $r=0.5$. We can see that the former exhibits dynamics qualitatively similar to Fig.~\ref{Fig4}(a) while the latter shows dynamics qualitatively similar to Fig.~\ref{Fig4}(b) in support of our analysis (note that $\gamma_{\max}$ and the initial $\rho$ values are not numerically identical to the values in the discrete model). 
The numerical solutions were obtained using the MATLAB package Chebfun~\cite{driscoll2014chebfun}.

\section{Conclusions}
Our simulations and models indicate that soft-spot overlap provides a mechanism that allows for entirely irreversible dynamics even in a system with a relatively small number of particles, such as an atomistic simulation. 
As the amplitude $\gamma_{\max}$ increases, the number of plastic events occurring during a cycle increases, and system-spanning avalanches become more frequent \cite{Regev2015}.
This increases the probability of having at least one new irregular overlap during each monotonic part of a forcing cycle, which, as we saw above, prevents the system from reaching the same zero strain state after a cycle. When this probability is large enough, the probability of the system returning to a zero strain configuration after a cycle vanishes, preventing it from ever reaching a periodic state. 
In future work, we plan to extract $p_{ir}$ directly from atomistic simulations and quantify the role played by avalanches in irregular overlaps and irreversibility. We would also like to clarify the relationship between our overlap approach, existing elastoplastic models, and strain compatibility \cite{jagla2007strain,tyukodi2018diffusion}. We believe that overlaps between plastic events may play a role in other mechanical systems, such as crumpled and corrugated sheets~\cite{Bense2021, Shohat2022}, and in the emergence of multi-periodic cycles as the irreversibility transition is approached~\cite{Szulc2022, Keim2021}. For these reasons, we believe that overlaps between plastic events will play an important role in understanding plasticity in mechanical systems.


\begin{acknowledgements}
{\it Acknowledgements}:
IR would like to thank Muhittin Mungan for useful discussions.
AS and IR were supported by the Israel Science Foundation (ISF) through Grant No. 1301/17 and AS was also supported by the Israel Science Foundation (ISF) Grant no. 2300/19.
\end{acknowledgements}
\bibliography{Bibliography}

\end{document}


\title{Supplementary material for: Overlapping plastic events as a mechanism for irreversible dynamics in amorphous solids under oscillatory shear
}
\maketitle
\date{\today}
\subsection{MD Simulation details}
We simulated 1024 particles in two dimensions, interacting with a radially-symmetric potential where the equations of motion were integrated with a leap-frog solver~\cite{Allen2017}. 
The potential that we used has a repulsive part identical to the Lennard-Jones potential and a minimum followed by a small hump that simulates the effect of bond-breaking~\cite{Lerner2009, Regev2013}. 
In our simulations, $50$\% of the particles were $1.4$ times larger than the other $50$\% to prevent crystallization. We prepared initial configurations by performing molecular dynamics simulations with an Andersen thermostat in which the system was kept at a high temperature $T=1$ (in Lennard-Jones energy units) in a liquid state. The simulation ran for $20$ simulation time units, after which the temperature was reduced to $T=0.1$ and was then run for another $50$ simulation time units. The final configuration was then quenched to zero temperature using the FIRE minimization algorithm that uses the molecular-dynamics integrator to find a nearby local minimum \cite{Bitzek2006}. 
This initial configuration was then subject to athermal quasistatic shear using the Lees-Edwards boundary conditions \cite{Lees1972}. This procedure was used to generate the different stable configurations and the plastic transitions between them as is described in \cite{Szulc2022}. Some of these configurations and transitions are reproduced in Fig. 1 of the main text.
\subsection{Model details}
In our simulation, there are $N$ sites, each capable of containing up to four active soft spots at any given moment. Each soft spot $a_i$ in site $i$ is modeled as a hysteretic element that can be in one of two states, denoted by $s[a_i] = 0$ or $s[a_i] = 1$. In a given site $i$, two of the soft spots are in state $0$ and two are in state $1$. The state of each soft spot switches when the stress in the site, denoted by $\sigma_i$, exceeds its stress thresholds. Specifically, a soft spot $a_i$ switches from $s[a_i] = 0$ to $s[a_i] = 1$ when $\sigma_i > \sigma^{+}[a_i]$, and switches from $s[a_i] = 1$ to $s[a_i] = 0$ when $\sigma_i < \sigma^{-}[a_i]$. Note that the stress $\sigma_i$ felt by all four soft spots in site $i$ is identical.
The stress thresholds of the different soft spots are drawn from a folded normal distribution, such that the stress threshold of soft spot $a_i$ in site $i$ is
\begin{equation}
	\sigma^{\pm}[a_i] = \pm\sigma_{th} \,|X|,
\end{equation}
where $X \sim {\mathcal{N}}(0,0.09)$, $\sigma_{th}=2$ and
\begin{equation}
	\sigma^{+}[a_i] >\sigma^{-}[a_i],
\end{equation}
which guarantees hysteretic behavior. After $\sigma^{+}[a_i]$ is reached, the stress drops by $\sigma^{drp}[a_i]$, which is a random number drawn from a uniform distribution ${\mathcal{U}}[0,1)$ that is decreased (for $s[a_i] = 0 \to 1$) or increased (for $s[a_i] = 1 \to 0$) from $\sigma_i$ of the corresponding site stress. To prevent infinite loops where the stress drop is large enough to cause the soft spot to switch back and forth, $\sigma^{drp}[a_i]$ is defined as
\begin{equation}
	\sigma^{drp}[a_i] = 0.5 (\sigma^+[a_i] - \sigma^-[a_i])\,Y,
\end{equation}
where $Y \sim {\mathcal{U}}[0,1)$.

Our simulation is event-driven, and thus at each simulation step, the algorithm searches for the smallest value of 
  \begin{equation}
    \delta \sigma =
    \begin{cases}
      \sigma^+_{(i)} - \sigma_i, & \text{for increasing strain and}\ s_{(i)}=0, \\
      \sigma_i - \sigma^-_{(i)}, & \text{for decreasing strain and}\ s_{(i)}=1,
    \end{cases}
  \end{equation}
where the subscript $(i)$ refers to all the active soft spots in site $i$. The soft spot corresponding to the lowest $\delta\sigma$ switches, and the strain in all the sites is then increased (or decreased) by $\delta \gamma = \delta \sigma / \mu$, where the shear modulus $\mu$ has been chosen to be one. 
When a soft spot at site $j$ switches, the stress at the rest of the sites is changed according to
\begin{equation}
\sigma_i = \sigma_i + G_{i,j},
\end{equation}
where
\begin{equation}
G_{i,j} = \frac{\eta_{ij}}{\sqrt{N}} + \frac{\tilde\eta_i}{N}
\end{equation}
is an interaction kernel. Here $\eta_{ij} \sim \mathcal{U}[-0.2,0.2)$, and $\tilde\eta$ is chosen such that
\begin{equation}
\sum_{j\neq i}G_{i,j} = \sum_{j\neq i}\frac{\eta_{ij}}{\sqrt{N}} + \tilde\eta_i = 0,
\end{equation}
\begin{equation}
\tilde\eta_i = -\sum_{j\neq i}\frac{\eta_{ij}}{\sqrt{N}},
\end{equation}
which is consistent with the conservation of stress in the quadrupolar interactions generated by soft spots, as was discussed in~\cite{Lin2014, Bocquet2009}.
%
\begin{figure}[tb]
\includegraphics[width=\linewidth]{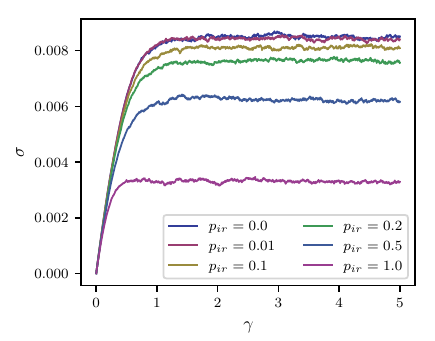}
\caption{Stress-strain curves for a system of $N = 250$ sites, averaged over $100$ realizations, for different $p_{ir}$ values. We used a wide stress distribution characteristic of poorly-annealed samples.
}  
\label{Sup_Fig1}
\end{figure}
Similarly to what was shown for monotonically increasing strain \cite{Kumar2022, Liu2022}, in many cases, when a site switches, the release of stress to the rest of the material causes other sites to become activated, which triggers an activation cascade or avalanche. To take this into account, after each switching event, we assess the instability $\Delta\mathcal{E}$ of the active soft spots by
\begin{equation}
	\Delta\mathcal{E} =
\begin{cases}
	|\sigma_i - \sigma^+_{(i)}|,& \text{if } s_{(i)}=0 \text{ and } \sigma^+_{(i)} < \sigma_i\,, \\
    |\sigma_i - \sigma^-_{(i)}|,&  \text{if } s_{(i)}=1 \text{ and }\sigma^-_{(i)} > \sigma_i\,. \\
\end{cases}
\end{equation}
We then switch the soft spot with the larger instability and repeat the process until the system is stable.

\subsection{Behavior under monotonically increasing strain}
We checked the response to a monotonically increasing strain for a system of $N=250$ sites, where the probability of irregular overlaps $p_{ir}$  was controlled. In Fig.~\ref{Sup_Fig1},  we show the stress-strain curves obtained for different $p_{ir}$ values, averaged over 100 realizations. These curves are characteristic of poorly-annealed samples. Different values of $p_{ir}$ were evaluated and found to lower both the maximum stress and the yield point. However, the stress-strain curves for $p_{ir}=0$ and $p_{ir}=0.01$, the values used in the simulations discussed in the main text, were almost identical. 

\begin{figure}[tb]
\includegraphics[width=\linewidth]{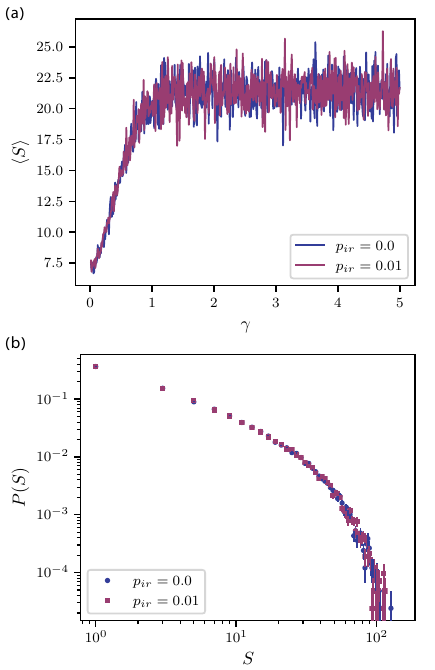}
\caption{Avalanche distribution for a system with $n=250$ sites. (a) Mean avalanche size versus strain averaged over $500$ realizations for $p_{ir}=0$ and $p_{ir}=0.01$. (b) The probability distribution of the avalanche sizes for post-yield strains averaged over $300$ realizations, exhibiting a power-law behavior with an exponential cutoff.
}  
\label{Sup_Fig2}
\end{figure}

\subsection{Avalanches}
Avalanche statistics for similar models was studied in the past \cite{Lin2014,lin2014scaling}. Here our goal was to verify that a model that includes overlaps reproduces similar results. For that purpose, we calculated the average avalanche size for two overlap probabilities, $p_{ir}=0$ and $p_{ir}=0.01$, with N = 250. Fig.\ref{Sup_Fig2}(a) compares the mean avalanche size, calculated from 500 realizations, and shows that in both cases it changes in a similar manner with respect to the strain. In Fig.\ref{Sup_Fig2}(b) we show the probability distribution of avalanche sizes taken from post-yield configurations, averaged over 300 realizations. This distribution follows a power-law with an exponential cutoff as was observed in related modesl ~\cite{Andresen2013, Sethna2001, Dahmen2009, Bhaumik2022}. We see that the post-yield avalanche size distribution also does not depend on $p_{ir}$, or at least depends very weakly on it.

\bibliography{Bibliography}